# Thermal stability and structural ordering of thick $Co_2MnAl$ films grown on MgO and Si substrates

David Robles-Cuenca[1], Juan Antonio González[1], Juan Pedro Andrés[1], Miguel Ángel Arranz[1], Mohamed Salaheldeen[2,3,4], Ricardo López Antón[1*]

[1] Regional Institute for Applied Scientific Research (IRICA) and Department of Physics, University of Castilla-La Mancha, 13071, Ciudad Real, Spain

[2] Dept. of Polym. and Adv. Mater., Univ. Basque Country, UPV/EHU, 20018 San Sebastian, Spain

[3] Department of Applied Physics I, EIG, University of the Basque Country, UPV/EHU, 20018 San Sebastián, Spain

[4] Physics Department, Faculty of Science, Sohag University, Sohag 82524, Egypt

* Correspondence: ricardo.lopez@uclm.es

Abstract: While most studies on $Co_2MnAl$ (CMA) focus on the ultrathin regime for spintronic junctions, applications scaling with magnetic volume—such as Anomalous Nernst Effect (ANE) energy harvesters, integrated power inductors, and magnetic MEMS—require significantly thicker films. In this work, 750 nm thick CMA films were grown on MgO (001) and Si substrates to establish a roadmap for their thermal integration. Structural analysis reveals that B2 ordering is optimized on MgO at 600 °C, which is important for functional devices. In contrast, samples on Si substrates exhibit a disorder transition toward the A2 phase, closely linked to diffusion-driven degradation processes. Above 500 °C, Si interdiffusion into the Heusler layer triggers the formation of interfacial silicides and Kirkendall voids, leading to microcracking, electrical resistivity jumps, and a final fully disordered A2 phase. Despite the Mn-poor stoichiometry ($Co_{2.25}Mn_{0.7}Al_{1.05}$), the films maintain a robust saturation magnetization of ~600 emu/cm³, consistent with recent theoretical models for off-stoichiometric Heusler alloys. These results define the processing windows for integrating thick CMA layers into multifunctional CMOS-compatible platforms.

## 1. Introduction

Heusler alloys have attracted considerable attention over the past decades due to their remarkable electronic and magnetic properties, which make them promising candidates for spintronic applications [1,2]. Among them, Co-based full-Heusler alloys with the general formula $X_2YZ$ (where X and Y are transition metals and Z is a main-group element) have been extensively studied because of their theoretically predicted half-metallic character and high Curie temperatures, with spin polarizations higher than 85% the Fermi level having been reported [3]. Such a property is essential for enhancing the performance of spintronic devices such as magnetic tunnel junctions (MTJs), magnetic random-access memories (MRAM), and spin valves.

In particular, $Co_2MnAl$ (hereafter CMA) has emerged as an attractive member of the Co-based Heusler family. Full Heusler alloys crystallize into a cubic structure but exhibit varying degrees of disorder. The highest order level is $L2_1$, which is where the highest spin polarization can be found in most alloys, but it is very difficult to achieve this order level. CMA, however, has a high spin polarization (76%) in its B2 structure, which is easier to achieve, according to the results of Kubota et al [4]. In that sense, recent articles from Kumar et al, focusing on the effect of the B2 disorder [5] and point defects [6] on the electronic and magnetic properties, confirm the high spin polarization of the B2 structure, with values close and sometimes even higher than in the $L2_1$ structure. Noteworthily, according to those DFT simulations, the magnetic properties are almost not affected by the disorder. Therefore, CMA is especially well-suited for thin-film applications where perfect atomic ordering is challenging.

As commented before, the $L2_1$ structure of CMA is fairly demanding to obtain by magnetron sputtering. In fact, only the group of Liu et al [7] has claimed to achieve it by epitaxial growth on GaAs substrates using magnetron DC sputtering and a post-annealing at 600 ºC. It is worth nothing that in order to determine that in their case it was the $L2_1$ structure which was obtained, Transmission electron microscopy (TEM) diffraction measurements were required, as the X-ray diffraction (XRD) measurements did not show several of the reflections of that structure due to the epitaxial growth. Meanwhile, several groups have managed to obtain B2 structure in $Co_2MnAl$ by sputtering growth on different substrates (Si, MgO) and/or buffer layers (Cr, Ag, Cu), either growing at a certain substrate temperature ($T_S$) [4,8–10] or applying a post-annealing after growing at room temperature [11–14]. Noteworthily, several groups have focused on the use of these films for magneto-transport applications. Hence, the group of Kubota et al, looking for adequate materials for tunnel junctions, grew CMA thin films on $Si/SiO_2$ and $Si/SiO_2/Cr$ substrates at different substrate temperatures (up to 600 ºC) [4] but also on MgO/Cr substates at room temperature followed by a post-annealing at 300 ºC [11]. Qiu et al [12] also prepared tunnel junctions on top of a MgO buffer, depositing CMA at RT (with different Ar pressures and, therefore, slightly different compositions) followed by an annealing at 400 ºC, whereas Yilgin et al analyzed the temperature dependence of the resistivity of CMA films deposited on $SiO_2$ at RT and annealed at 200 or 400 ºC [13]. Sato et al [14] studied the anisotropic magnetoresistance of CMA deposited on a MgO substrate at RT and annealed one hour at 650 ºC and the group of Wang [10] analyzed the domain wall resistance of CMA thin films grown on MgO substrates at 500 ºC. Finally, Barwal et al [9] analyzed in detail the dependence of the magnetization reversal in CMA films grown at different temperatures (up to 500 ºC) on Si substrates.

While most studies on Heusler alloys focus on the ultrathin regime (< 50 nm) mainly for spintronic applications, the 750 nm thickness investigated in this work is strategically chosen for also studying its use in technologies that scale with magnetic volume. Specifically, such thicknesses are essential for Anomalous Nernst Effect (ANE) energy harvesters, where a substantial thermal gradient must be established across the film to maximize voltage generation. The 750 nm thickness provides a higher thermal resistance compared to ultrathin films, a structural prerequisite for establishing the temperature gradients necessary in thermal harvesting technologies. Furthermore, in integrated power inductors, thick layers are required to handle high magnetic flux without saturation. In the field of magnetic microelectromechanical systems (MEMS), the actuation force for components like micro-mirrors or valves depends directly on the magnetic volume and the saturation magnetization ($M_s$) of the active layer. Although other Heusler compounds may exhibit higher $M_s$ values, CMA stands out as a prominent Weyl semimetal[15], offering a unique platform for multifunctional devices. Consequently, this study provides a critical roadmap for the thermal integration of thick CMA films onto Si and MgO substrates, defining the processing windows necessary for these emerging volume-dependent applications.

Some of the possible applications of CMA require thermal stability at high temperatures. Therefore, we wanted to check the effect of the temperature during deposition on two different substrates (MgO and Si), reaching higher temperatures than usual (up to 700 ºC) and on relatively thicker films. In particular, we have studied the effect of the growth temperature of $Co_2MnAl$ thin films on the microstructure and topology of the films and, consequently, on their transport and magnetic properties. In that sense, $Co_2MnAl$ thin films (with thickness about 700 nm) were grown on MgO and $Si/SiO_2$ (i.e., without removing the naturally oxidized layer) substrates, from RT up to 700 ºC, comparing also the effect of the heating during deposition with a post-annealing of a sample grown at RT.

## 2. Experimental details

$Co_2MnAl$ thin films have been obtained in a radio frequency (RF) magnetron sputtering system with a base pressure of $10^{-7}$ mbar. The Ar pressure during the deposition was $6.5 \cdot 10^{-3}$ mbar (with an Ar flow of 25 sccm). Films were grown from a $Co_2MnAl$ target (supplied by PhotonExport). However, due to the difference in sputtering yield in each element, the composition of the obtained films was $Co_{2.25}Mn_{0.7}Al_{1.05}$, according to Energy Dispersive Spectroscopy (EDS) measurements. The power of the RF source was kept constant at 50 W. The growth rate (approximately 4.2 Å/s) was determined by x-ray reflectometry performed on few-nanometers thin films deposited on monocrystalline silicon under the same growth conditions. The main measurements were carried out on relatively thick samples grown for half an hour, giving a nominal thickness of 750 nm, which was later confirmed by Scanning Electron Microscopy (SEM) images. The samples were deposited on a temperature controlled heated substrate holder in front of the target (at a source-to-substrate distance of 4.8 cm). The samples were grown at the following temperatures: 25ºC (room temperature), 300ºC, 400ºC, 500ºC, 550ºC, 600ºC, and 700ºC. The substrates on which the films were grown were silicon wafers with orientation (001) and MgO squares with orientation (001). In order to determine the best way to grow samples, we compared the properties of samples grown at high temperature with others grown at room temperature and then annealed, both kinds of samples being grown in the same system under the same conditions. For this purpose, samples grown at room temperature were also annealed at 500ºC and 700ºC, and it was observed that in the case of those annealed at 700ºC on silicon, the $Co_2MnAl$ layer was delaminated due to the mismatch in thermal expansion coefficient between the substrate and the film. Therefore, no comparison is possible with the samples grown directly at high temperature for Si substrate.

The composition of the samples was obtained by EDS measurements performed in a scanning electron microscope (Zeiss GeminiSEM 500, Oberkrochen). X-ray reflectivity and diffraction (in 2θ geometry) measurements were performed in a Bruker D8 Advance system (Bruker, Billerica), using CuKα radiation (λ = 1.54 Å). Hysteresis loops at room temperature were measured using a Vibrating Sample Magnetometer (VSM model 9500, LDJ). An Atomic Force Microscope (AFM) by NT-MDT was used in tapping mode at room temperature for studying the topography of the films. High-resolution AFM probes (SHR150, BudgetSensors) were used for the AFM measurements. The same microscope was configured to also operate with two-pass modes for carrying out Kelvin Probe Force Microscopy (KPFM). In this technique, the conductive tip ($W_2C$, NT-MDT) scans the surface sample (xy plane) and locally measures its work function, hence obtaining a xy-map of the surface electric potential. As a two-pass mode, a preliminary topographic image is recorded and stored. Later, the conductive tip is lifted a vertical distance (Dz) to proceed with the mapping of the surface electric potential, following the stored topographic profiles. For this second pass, Dz is appropriately selected to minimize the influence of the topographic features, the resulting measurement being mediated only by the long-range electric interaction. Consequently, this KPFM gives information about the electronic state of any local structure on the sample surface. For instance, KPFM is particularly useful to detect charge trapping in dielectrics or anomalies of the surface electrical potential of metals. Resistivity was measured using a custom-built measurement setup implementing the Van der Pauw method with spring-loaded probe pins.

## 3. Results and discussion

### 3.1. Structural characterization (XRD)

Before showing our XRD results, we would like to briefly summarize the different kinds of ordered phases found in the full Heusler alloys $X_2YZ$ versus the characteristic peaks expected in their XRD patterns. Full Heusler alloys such as $Co_2MnAl$ can crystallize in the fully ordered $L2_1$ structure, which consists of four interpenetrating FCC sublattices. When Mn and Al atoms exchange positions randomly, a partial disorder known as the B2 structure is formed. If Co atoms partially swap positions with Mn atoms with equal probability, the resulting configuration is referred to as $DO_3$-type disorder. Complete random occupation of X, Y, and Z sites by Co, Mn, and Al leads to XYZ disorder, corresponding to the fully disordered A2 phase. It should be noted that $DO_3$ is considered a variant of A2 disorder and cannot be clearly identified by XRD because Co and transition metals at Y sites have similar scattering factors. The identification of the structures through XRD relies on the observation of the specific peaks corresponding to each phase. The A2 structure exhibits reflections where h, k, and l are all even and the sum (h + k + l) is a multiple of 4. In the presence of B2 ordering, additional peaks appear compared to A2, as the condition for non-zero structure factor amplitude requires only that h, k, and l are even. For the perfectly ordered $L2_1$ structure, extra reflections occur for planes with Miller indices that are either all odd or all even [9]. Consequently, the observation of (111) and (311) peaks indicates $L2_1$ ordering, whereas their absence combined with the presence of (200) suggests B2 ordering. Finally, the lack of (111), (311), and (200) peaks implies the formation of the fully disordered A2 phase.

The X-ray diffractograms of all samples can be seen in Figure 1. Focusing on the films grown on MgO (see Figure 1 (a)), an increase in the intensity of the peaks at a sample angle of 33º (200) and at 65º (400) can be observed as we increase the substrate temperature, reaching their maximum intensity around 600ºC. This clearly indicates the presence of a B2 structure. The (220) peak at 44º usually appears as well, but since the substrate MgO peak is located at that position, it is not possible to verify its presence.

In the samples grown on silicon, similarly to what occurred previously with the MgO peak, the presence of the (400) peak cannot be verified due to the (004) silicon peak being located at the same position. Meanwhile, the (200) peak does not appear at any temperature, whereas the (220) peak appears with its highest intensity at room temperature and decreases as the temperature increases, reaching an almost negligible amplitude at 500ºC, but increasing afterwards for 600ºC and 700ºC. Note that for 700ºC there is also another small peak at 46.5º, which should not appear for the CMA. As the peak (220) is observed but not the peak (200) or (400), the structure must be A2 in all the samples grown on silicon. We remark that in our case, we obtain the B2 structure in the samples grown on MgO substrates at temperatures higher that RT, agreeing well with the results found by, e.g., Sato[14] and Wang[10] but, nevertheless, the structure we obtained for the samples grown on Si was always A2. This is in contrast to the case of Barwal[9] which also found the (400) reflection for samples grown at 500º C on Si substrates (although with fairly low intensity) and considered that the phase was quite likely B2 (with an intensity of the (200) reflection below the detection sensitivity of their XRD system) or the case of Kubota[4], who also found the (200) reflection of the B2 structure for samples grown at 300 ºC. This could be related to the fact that the composition of our samples was not exactly the stoichiometric one.

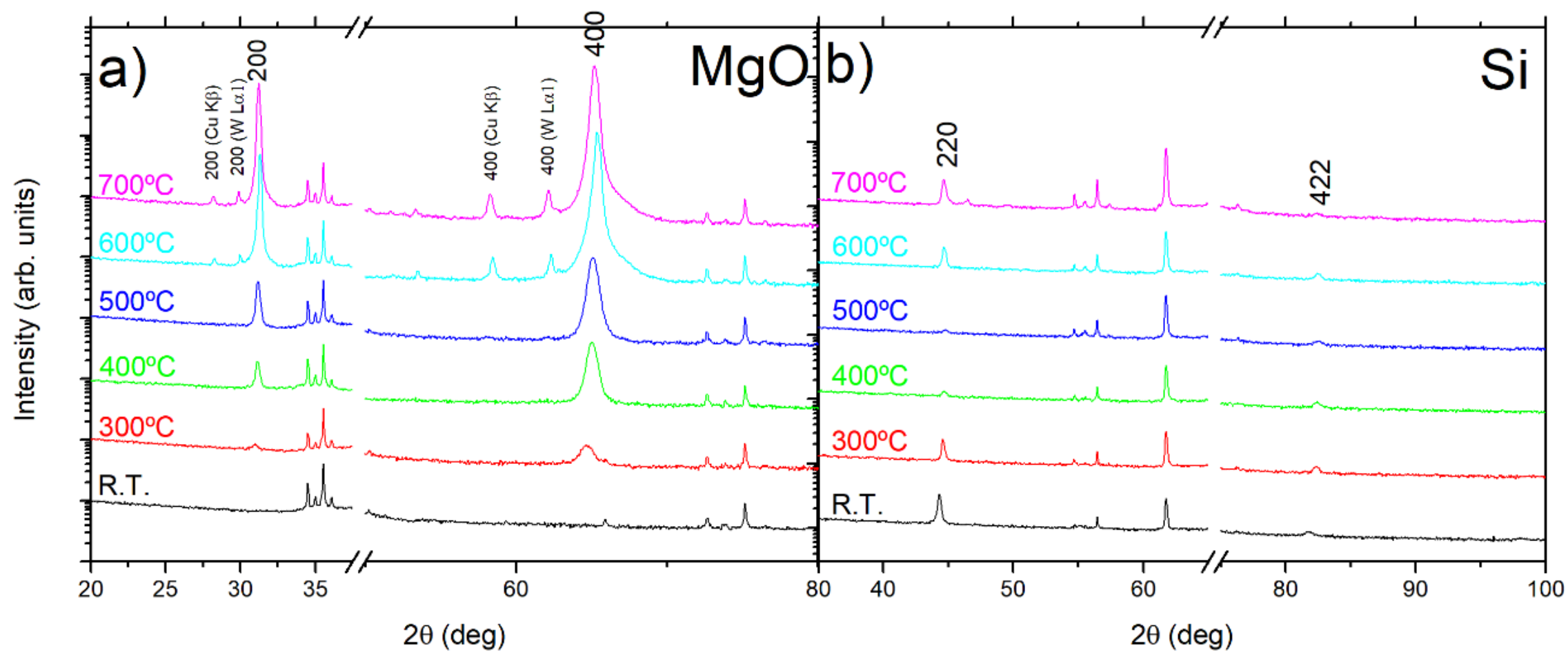

Figure 1. X-ray diffraction measurements in $Co_2MnAl$ grown at different temperatures on (a) MgO and (b) monocrystalline silicon. Unlabeled peaks correspond to MgO and Si substrates, respectively.

From the X-Ray Diffraction measurements, the lattice parameter and grain size were calculated for all the samples. Grain size was obtained through Debye Scherer's formula:

$$D_{hkl} = \frac{K\lambda}{\beta_{hkl}\cos\theta}$$

Where $D_{hkl}$ is the gran size in the perpendicular direction of the sample, hkl are the Miller indices of the planes, K is the crystallite-shape factor (taken usually as 0.9), $\lambda$ is the wavelength of the X-rays, $\beta_{hkl}$ is the width of the peak in radians and $\theta$ is the sample angle.

For samples grown on MgO, it can be seen in Figure 2 a) how growing at high temperature the lattice parameter (with its theoretical value of 5.75 Å indicated by a line) decreases, with the lattice being compressively strained by 0.7%, and the crystallinity improves by a factor of three. The grain size was higher than the obtained in Vilanova[8], which was of 20 nm for deposition temperatures between 450 and 700ºC. For the samples grown on Si there was no clear trend of grain size, while the lattice parameter appears to reduce from 5.775 Å when grown at room temperature to around 5.73 Å when grown at high temperatures. Meanwhile, the lattice parameters are similar to those obtained by Barwal[9], but, however, the grain size is four times higher than theirs. This could be related to the fairly thinner thickness of their films, restricting large grain sizes.

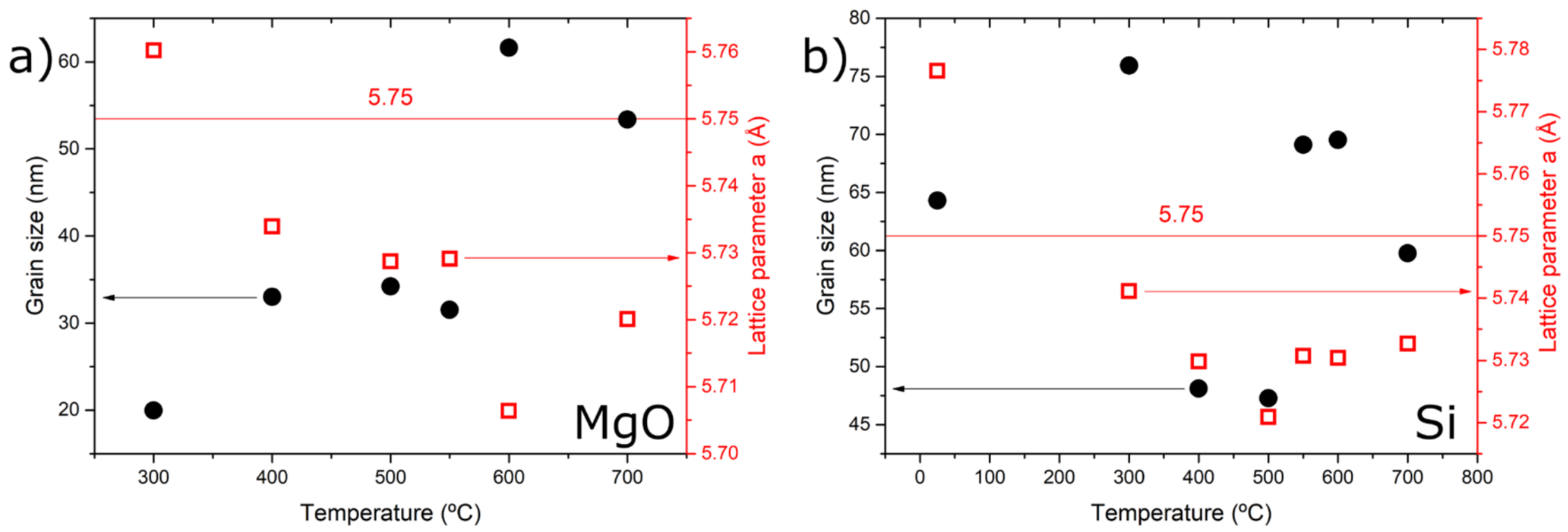

Figure 2. Grain size (left-black solid circles) and lattice parameter (right-red empty squares) against growth temperature of Co2MnAl in a) MgO and b) Si. The dotted line indicates the theoretical value of the lattice parameter.

In summary, the XRD characterization demonstrates that a highly ordered B2 phase is successfully achieved on MgO substrates at a growth temperature of 600 °C. While the $L2_1$ structure represents the fully ordered state, the B2 phase is sufficient for most practical applications, as it has been shown that in a similar system ($Co_2MnGa$) B2 retains high spin polarization[11] and the characteristic Weyl semimetal behavior[16]. From a technological perspective, targeting B2 ordering significantly relaxes the thermal budget and reduces fabrication complexity, facilitating the cost-effective integration of CMA into functional devices without compromising its fundamental magnetic and electronic performance. Regarding spin-dependent applications, Si substrates are not a viable option; the observed A2 disorder collapses the half-metallic gap, preventing the material from reaching the high spin-polarization values preserved in the B2-ordered structure.

## 3.2. Surface morphology (AFM and HRSEM)

To establish the fundamental relationship between growth conditions and the resulting physical properties in CMA films, a thorough morphological and surface characterization is required. Given that both the magnetic response and the topological features of Weyl semimetals are intrinsically linked to surface quality and structural integrity, it is essential to systematically evaluate the influence of substrate choice and deposition temperature. In this regard, Atomic Force Microscopy (AFM) and High-Resolution Scanning Electron Microscopy (HRSEM) were employed to determine the processing windows and the physical stability of the layers, providing the necessary framework to interpret their subsequent magnetic behavior.

The average roughness was measured by averaging in squares of 4x4 μm$^2$ and the results are shown in Table 1. The general trend is an increase of roughness with the substrate temperature, this change being much more noticeable in silicon from 600ºC and above.

Table 1. Average Roughness measured by AFM.

| Temperature (ºC) | R.T. | 300 | 400 | 500 | 550 | 600 | 700 |
|---|---|---|---|---|---|---|---|
| Ra (nm), Si | 0.25 | 1.7 | 3.85 | 1.56 | 2.2 | 5.85 | 9.07 |
| Ra (nm), MgO | 0.35 | 0.84 | 2.2 | 2.32 | 1.5 | 2.4 | 3 |

While doing these measurements a pattern of lines was observed by the optical microscope on the surface of samples grown on heated Si for temperatures above 500ºC. Figure 3 shows the pattern observed for each temperature. In the samples grown at 500ºC there were few lines, parallel to each other, separated by several hundred micrometers. At 550ºC the separation between lines continues on similar scales but the corners begin to curve. At 600ºC the lines began to close forming smaller and more irregular divisions on the surface, with sizes ranging between 100 and 200 μm wide. Finally, at 700°C many more of these lines appeared and the divisions became much smaller, about 15 μm wide.

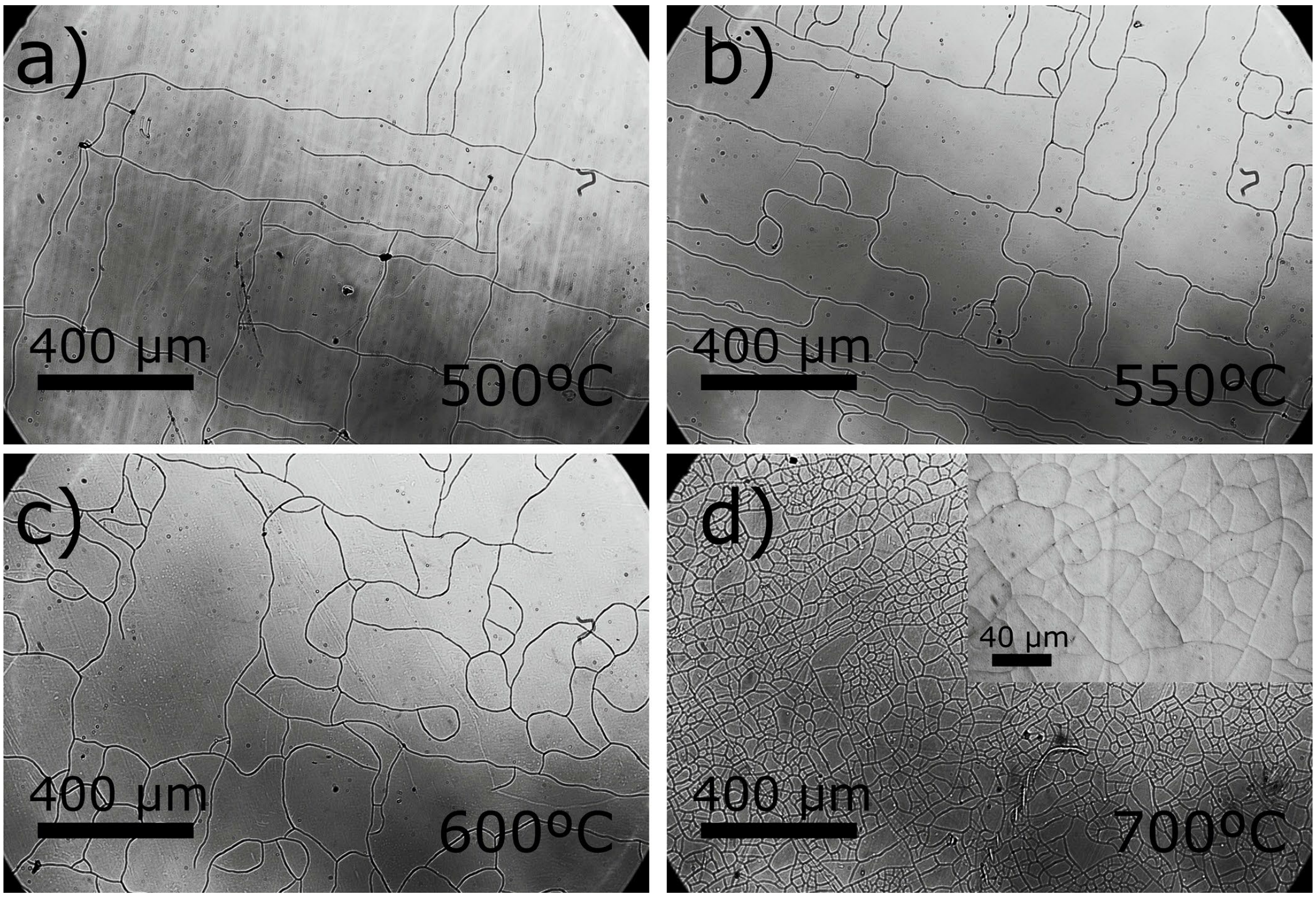

Figure 3. Optical microscopy images of the pattern of lines at the surface of CMA grown on Si at a) 500ºC, b) 550ºC, c) 600ºC and d) 700ºC. The 700ºC image has an inset with greater zoom.

Figure 4 a) shows an AFM measurement of 40x40 μm$^2$ made on a sample grown at 700ºC, where this pattern of lines can be observed, and, as seen in detail in the profile seen in Figure 4 b), the lines are small elevations on the surface. Such features were previously observed by optical microscope (see the panel of images in Figure 3) and now, we include here their characterization by KPFM technique. The results for that two-pass microscopy mode are shown in Figure 4. Beside the preliminary AFM topography (Figure 4 a)), the second-pass image concerns the mapping of the surface electric potential on the same scanned region (Figure 4 c)). Therein, the topographic contribution has been virtually removed upon lifting the tip by a Δz distance. Therefore, the Figure 4 c) would only show the surface anomalies (bright lines) responsible for altering the constant electric potential at the metallic sample (dark regions). In view of these KPFM results, we could explain these lines to be microcracks in the sample, where the electrical continuity notably changes with respect to the metallic regions, breaking the bidimensional symmetry of thar surface electric potential. Finally, the brightest contrast in KPMF signal, i.e. the strongest electric distortion, would correspond to the crack rising of those elevations, whose edges were already shown in Figure 4 b).

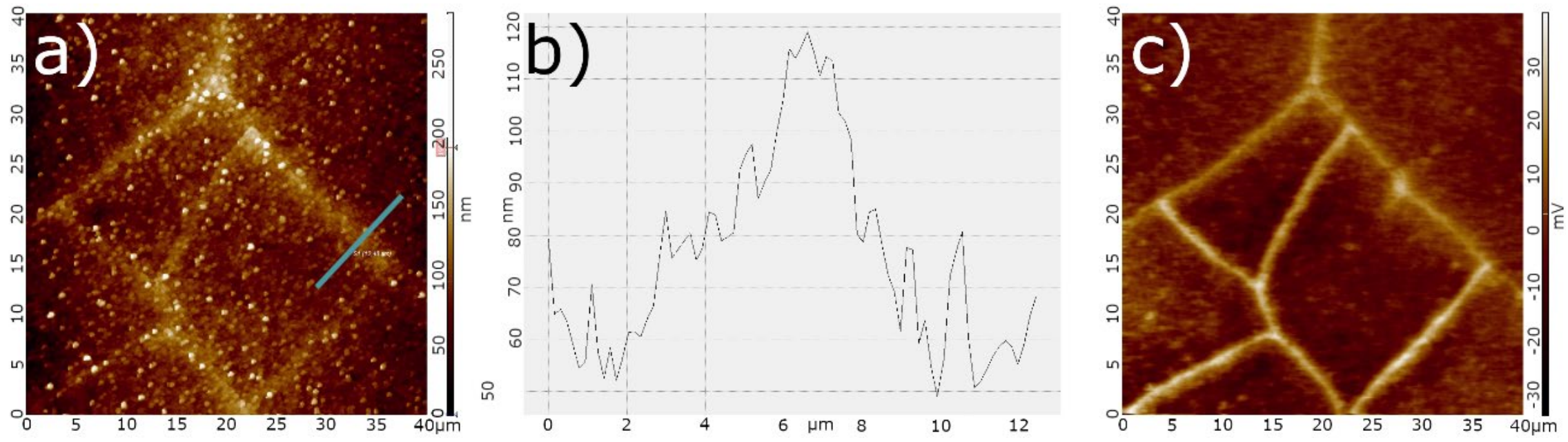

Figure 4. (a) AFM measurement for sample grown on Si at 700ºC at 40x40 µm scale, (b) profile of the blue line in the same measure and (c) KPFM of the same region.

We also checked the surface using SEM, as can be seen in Figure 5 a), where we clearly see in detail one of these microcracks in the sample grown on Si at 700 C. To obtain more information about these microcracks, we made a cross-sectional measurement of that sample, as it is the sample showing more of these structures. We could observe, as seen in Figure 5 b), that the microcracks entered even into the substrate and, even more, there were small voids in the silicon underneath the $Co_2MnAl$ layer. We checked the cross-sectional view of the rest of the samples grown in silicon, but only the 700ºC sample exhibits these voids. We performed a compositional analysis of the cross-sectional view for each sample, and we were able to observe that for samples grown at temperatures above 500ºC, there was a high concentration of Si in the lower part of the $Co_2MnAl$ layer, evidencing an important diffusion of Si. Figure 6 shows the profile for the samples grown at RT, at 500ºC and at 700ºC. The origin of the x-axis is located at the interface between the film (right) and the substrate (left). It should be noted that the compositional profile obtained is smeared out due to the finite size of the analysis probe. We will consider the case grown at room temperature as an example of what should be seen if there is no interdiffusion, since Si and CMA should not have been mixed in such conditions. We can see that at room temperature the composition of Co, Mn and Al in the film is stable and it rapidly decreases when entering the negative axis, as Si increases. In the case of the sample grown at 500ºC, the composition of Co, Mn and Al does not decrease so sharp as it did at room temperature, and the transition of the Si is also smoother. Finally, for the sample grown at 700ºC, it is highly evident the presence of Si up to 400 nm in the CMA layer, and the irregularity in composition of Co, Mn and Al through that region. This may indicate strong diffusion of Si into the sample when grown at temperatures above 500ºC. Co and Si usually react to form $CoSi_2$ when a film of Co on a silicon substrate is annealed at more than 500ºC[17], and the same reaction has also been reported in films of Co grown on Si at temperatures above 500ºC[18]. Therefore, we believe the same reaction is occurring in $Co_2MnAl$, being much more prominent at 700ºC. This could help to explain the magnetic behavior we find on the samples deposited on Si (discussed below), also being at the root of the formation of voids under the thin film while the CMA film maintains a remarkably planar morphology. This phenomenon is a textbook manifestation of the Kirkendall effect, where the outward diffusion of Si atoms into the magnetic layer occurs significantly faster than the inward diffusion of Co or Mn. The vacant lattice sites left behind by the migrating Si coalesce into macroscopic voids. The fact that the 750 nm film does not collapse into these cavities demonstrates its high mechanical rigidity and structural integrity, effectively 'bridging' the reaction zones and being another advantage of working with thick layers. This observation confirms that at 700 °C, the Si substrate acts as a reactive source rather than a passive template, leading to the formation of interfacial silicides as previously suggested by the XRD data.

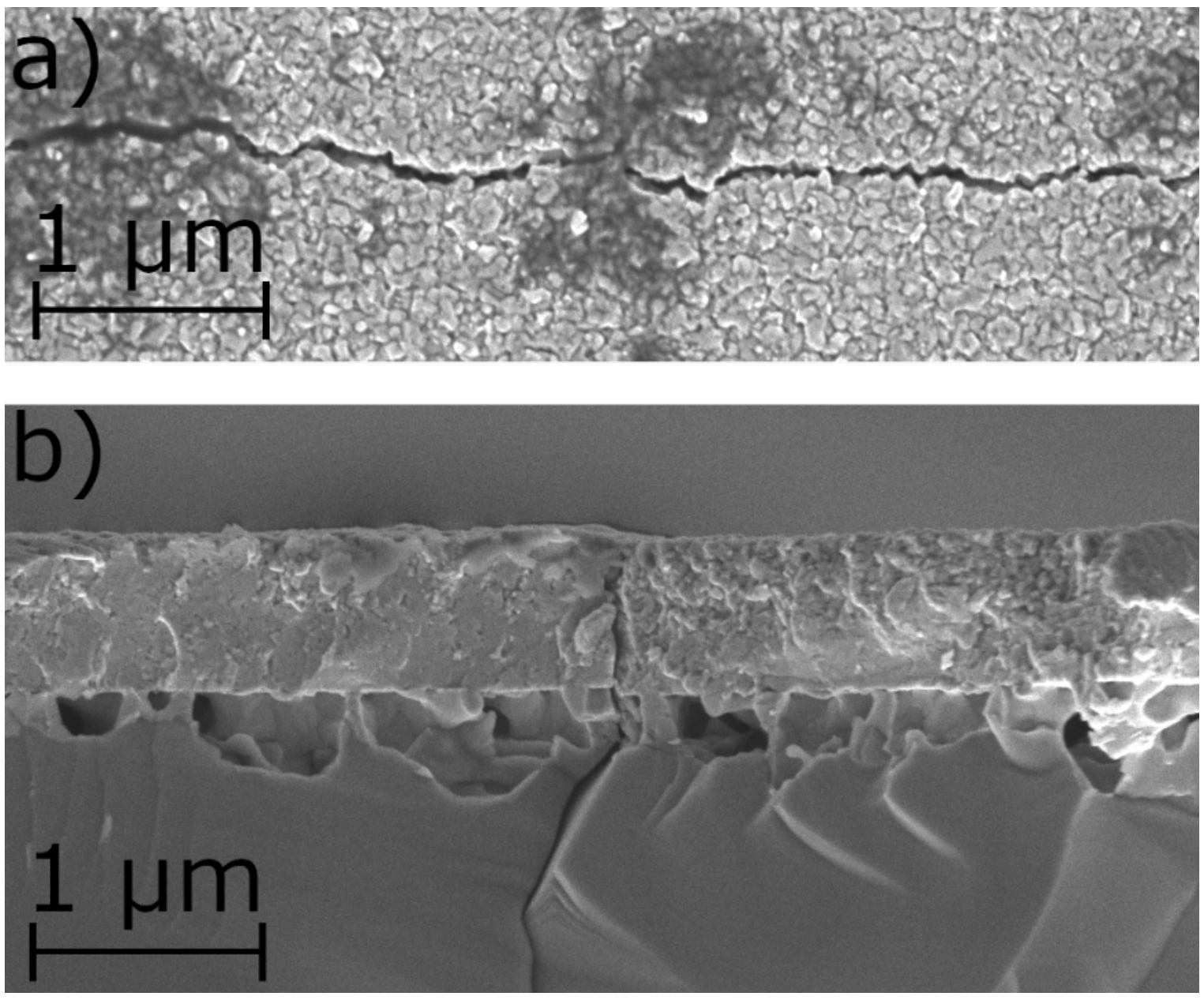

*Figure 5. a) One of the microcracks seen by SEM in a sample grown on Si at 700ºC and b) cross-sectional image of the same sample.*

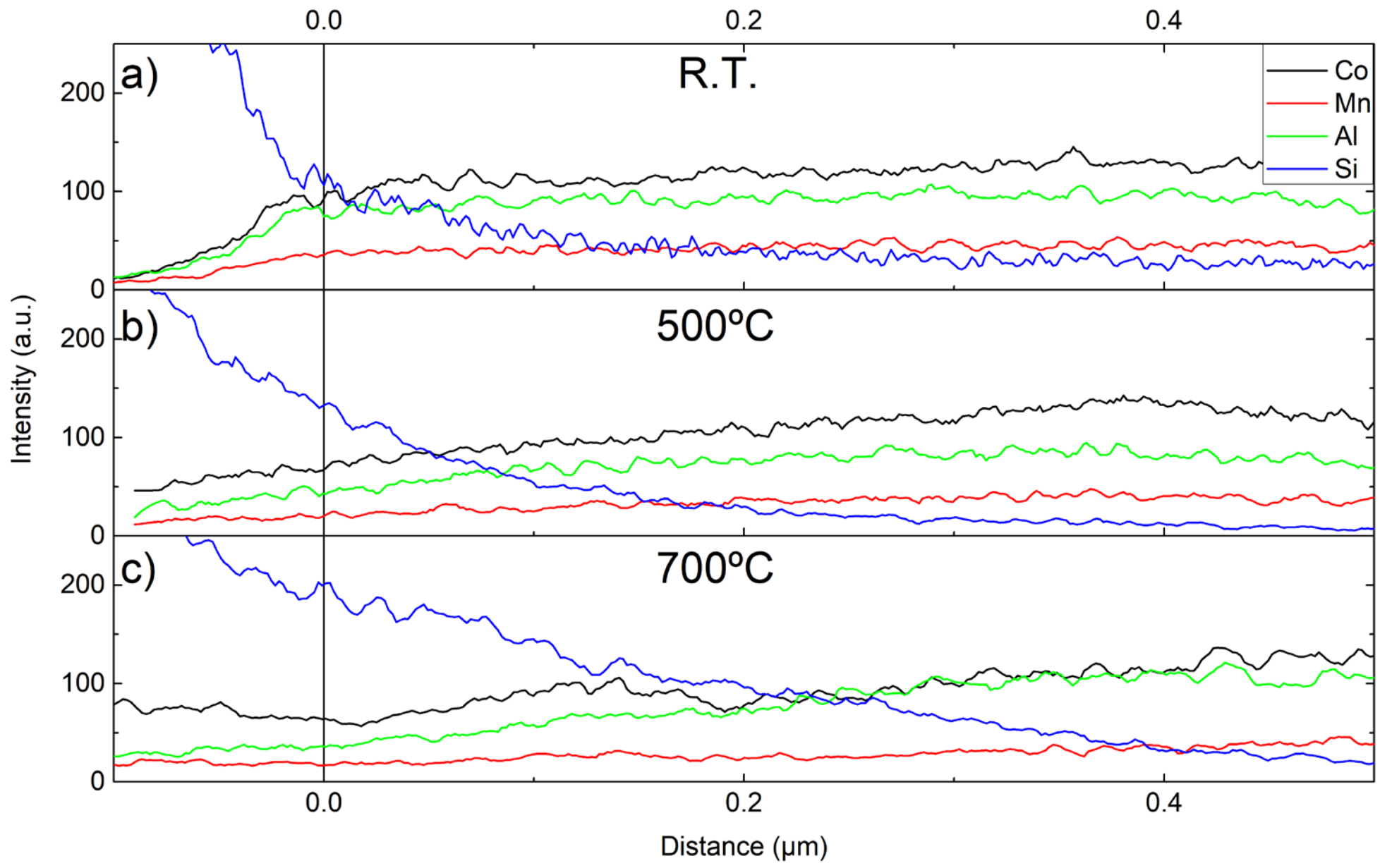

*Figure 6. Line composition in cross-sectional image of $Co_2MnAl$ grown on Si a) at RT, b) at 500ºC and c) at 700 °C.*

Altogether, these two techniques provide definitive evidence of the thermal degradation mechanisms occurring on the Silicon substrate. The visual identification of cracks, gaps, and diffusion processes at 700°C not only justifies the loss of structural order detected by XRD but also allows for the precise determination of the material's mechanical stability limits. Consequently, this morphological study acts as a fundamental diagnostic tool to define the technical viability window of CMA in complex devices, ensuring the physical integrity required for high-volume applications such as MEMS or integrated inductors.

### 3.3. Electrical properties.

Figure 7 shows the resistivity measurements in both types of substrates as a function of temperature. The values are very close to 250 µΩ·cm, typical values reported for $Co_2MnAl$ at room temperature in bulk[19], and also for thin films studied by Yilgin[20]. In the case of the samples grown on Si, the resistivity is around 260 µΩ·cm for the samples grown up to 400ºC. There is a jump in the resistivity at 500ºC, with a value around 310 µΩ·cm afterwards, and increasing even more at 700ºC. That increase in resistivity is likely related to the onset of Si diffusion into the layer, as well to the appearance of microcracks, which are much more numerous at the higher temperature. This is slightly different from what Yilgin observed in their films deposited on $SiO_2$: a clear increase in resistivity at 400ºC (although smaller than the one observed by us at 500ºC).

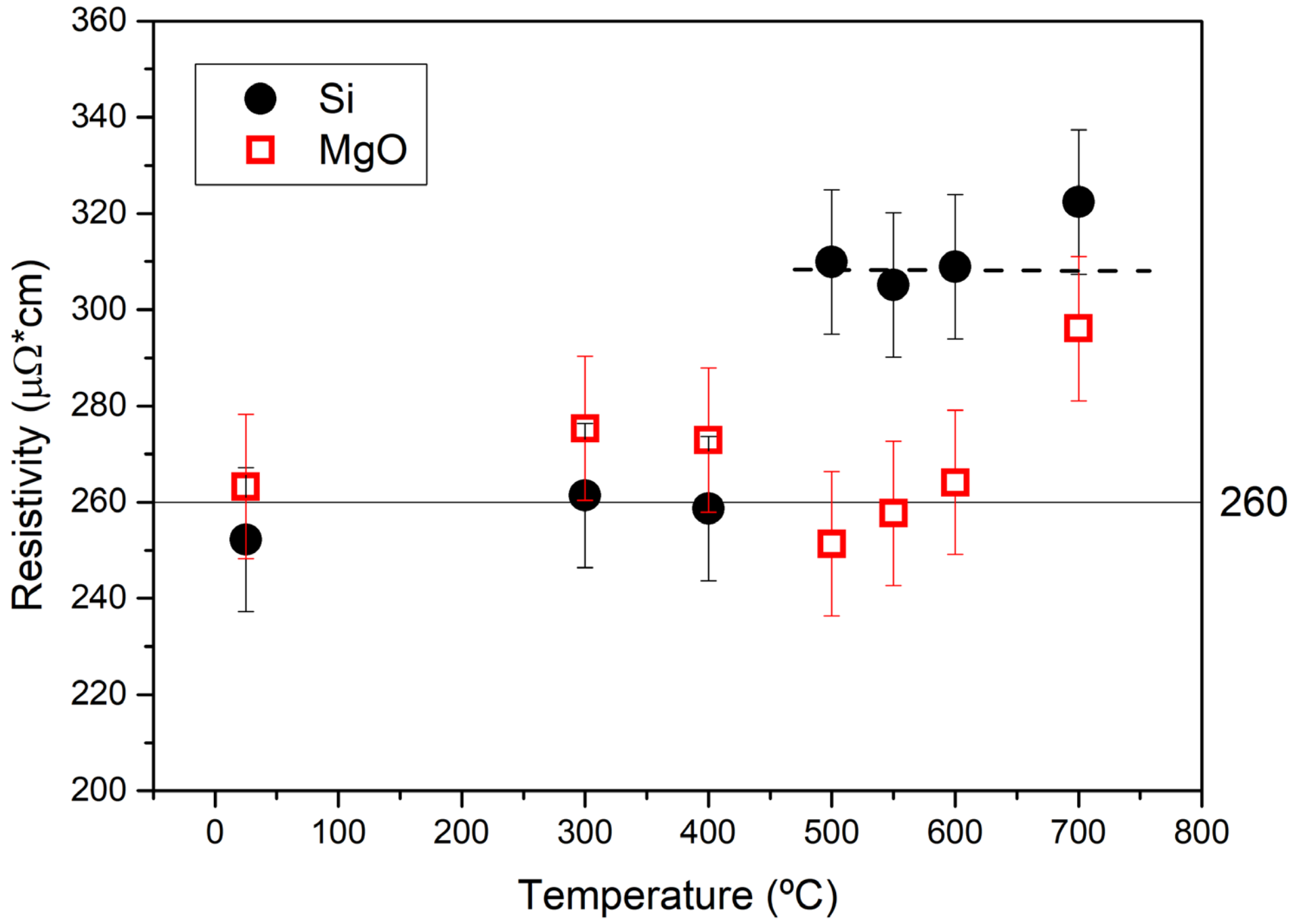

Figure 7. Resistivity as a function of temperature.

3.4. Magnetic properties.

Figure 8 a) and b) show, respectively, the magnetic hysteresis loops and the magnetic properties as a function of temperature measured for samples grown in MgO at R.T, 300ºC, 500ºC, 600ºC and 700ºC. It can be seen how the increase in the deposition temperature leads to a decrease in the coercive field and to an increase in the squareness of the loop. It is also noticeable how saturation magnetization increases with the deposition temperature until reaching its maximum at 600ºC and decreasing afterwards at 700ºC. In the case of Vilanova[8], its maximum saturation magnetization is for a deposition temperature of 522ºC, and magnetization is reduced by 35% at 688ºC. A similar behavior occurs in Yilgin [13] for annealed samples, where they have a maximum magnetization of 745 emu/cm$^3$ for 300ºC and decreases with increasing deposition temperature, remaining at 612 emu/cm$^3$ for 500ºC. The fact that our samples have their maximum at a higher temperature than in these two cases could be due to the difference in thickness, with our films being between 10 and 20 times thicker, which makes less relevant the influence of the substrate and/or seed layer. With regard to the coercive field, in the loop shown in Wang[10], a very small coercive field is observed for a sample grown at 500ºC, as in our case.

The experimental saturation magnetization of approximately 600 emu/cm$^3$ observed in our films is in excellent agreement with the compositional mapping recently reported by Sato[14]. In the CMA, Manganese is the primary source of the local magnetic moment (~3 μB/atom), while Cobalt provides a significantly lower contribution (~0.5-0.7 μB/atom) Consequently, the reduction of Mn content to y=0.7 in our samples leads to an intrinsic decrease in the total magnetic moment that cannot be fully compensated by the excess Cobalt (x=2.25), even if the latter occupies Mn antisite positions. This consistently explains the deviation from the ideal stoichiometric value of 4 μB/f.u. (~780 emu/cm3).

Remarkably, it has been shown that despite their off-stoichiometric nature, the fundamental magnetic characteristics of the CMA Weyl system are preserved[16].

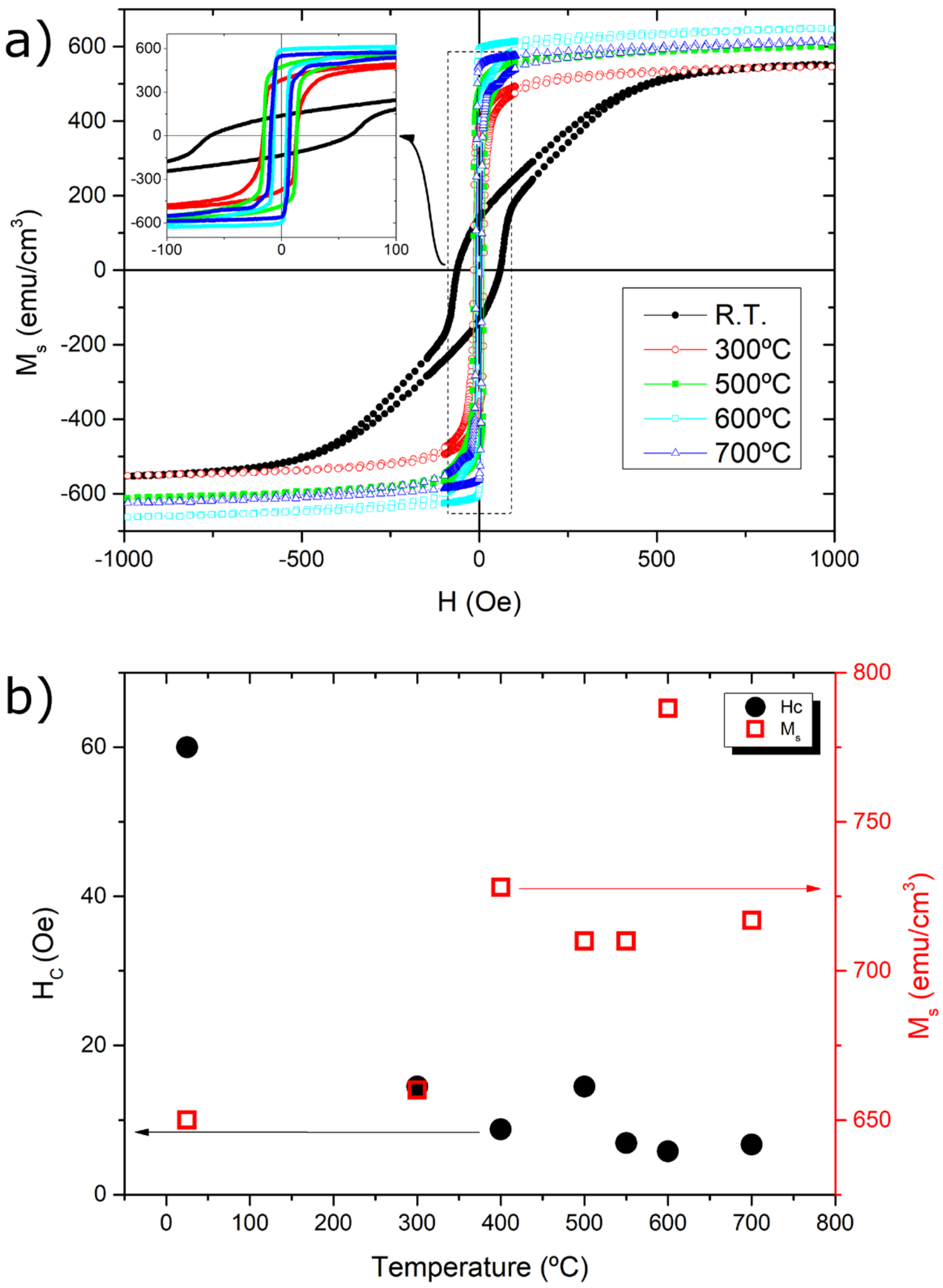

Figure 8. a) Hysteresis loops (inset: detail at low field) and b) coercive field (empty red squares) and saturation magnetization (full black circles) as a function of temperature for samples grown on MgO at different temperatures.

Figure 9 a) shows the magnetic measurements for samples grown on Si at: R.T, 300ºC, 500ºC, 600ºC and 700ºC. Figure 9 b) shows their magnetic properties as a function of temperature. In this case, we detect a similar behavior to that observed in samples on MgO: reduction of the coercive field and increase in saturation magnetization and squareness of the hysteresis loop, but only up to 500ºC. At higher temperatures, a rapid increase in the coercive field and a rapid reduction in saturation magnetization can be observed, which would be consistent with a change in composition given that Si is mixing with the alloy. In the case of Kubota[4] for CMA grown at high temperatures on Si, a maximum saturation magnetization of about 780 emu/cm$^3$ is reached for 300ºC, higher than in our samples. However, it also decreases as the growth temperature increases, being more pronounced at 600ºC, where they have a value similar to ours (420 emu/cm$^3$). This could indicate that similar diffusion processes might also be taking place in their samples. Kubota does not show the evolution of the coercivity but, nevertheless, their value at 300ºC is very similar to that of our samples. Meanwhile, Barwal[9] observe a behavior slightly different from ours, with the coercive field increases with deposition temperature, reaching its maximum at 350ºC, and then decreases for temperatures above 350ºC.

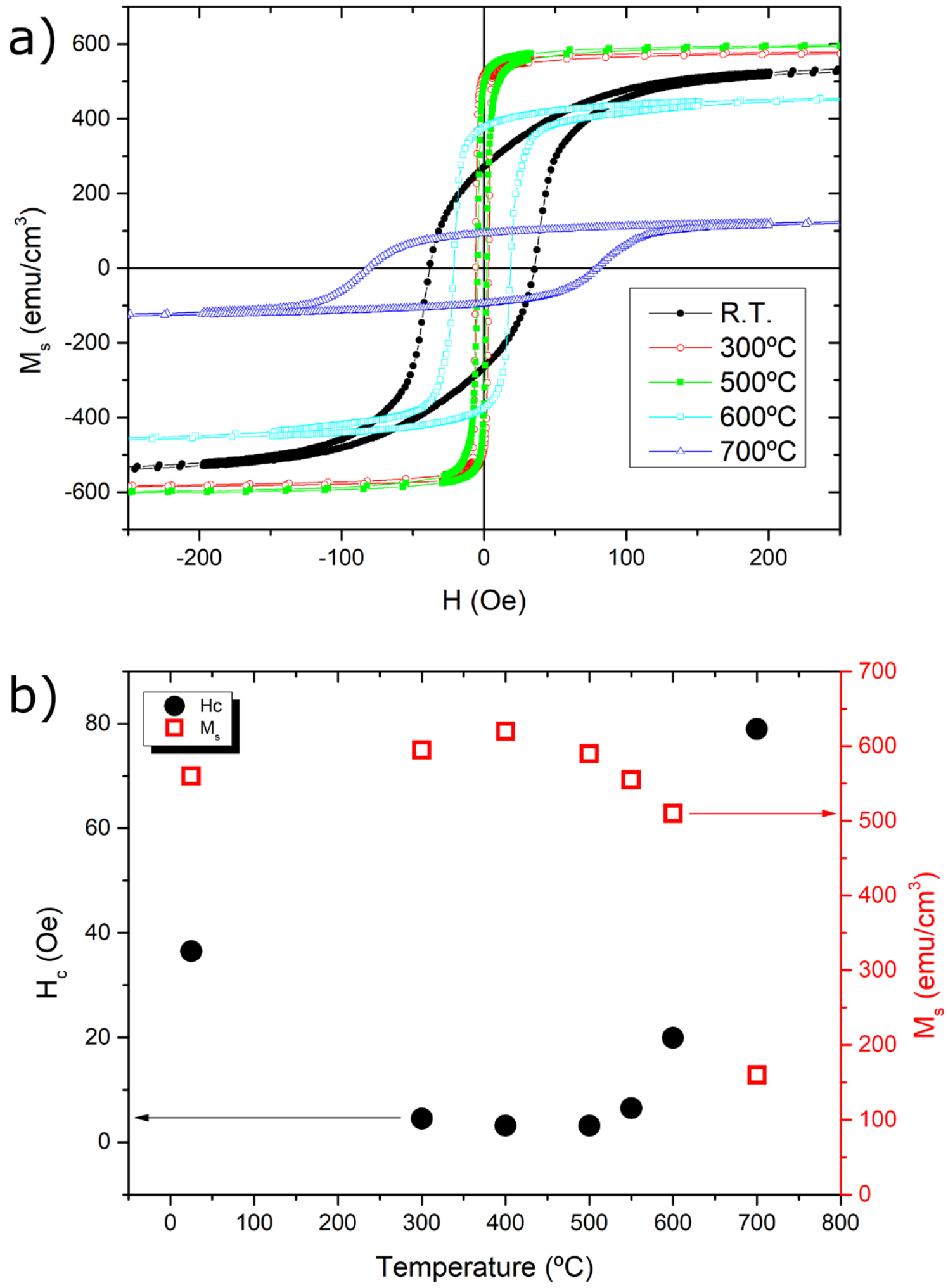

Figure 9. a) Hysteresis loops and b) coercive field (empty red squares) and saturation magnetization (full black circles) as a function of temperature for samples grown on Si at different temperatures.

It is noteworthy that despite the prevalence of the A2 phase and the structural degradation observed on Si substrates at high temperatures, the saturation magnetization remains remarkably consistent with the values obtained for B2-ordered films on MgO. This structural-magnetic decoupling indicates that the macroscopic magnetic response is relatively insensitive to the site-swapping disorder inherent in the A2 phase. Consequently, Si substrates emerge as a cost-effective and technologically viable platform for applications primarily requiring high magnetic flux or volume-dependent forces, such as integrated power inductors or magnetic MEMS. In contrast, for spintronic devices where high spin polarization or half-metallicity is mandatory, the Si-grown films remain unsuitable due to the detrimental effect of A2 disorder on the spin-resolved density of states. Therefore, the choice of substrate must be carefully balanced according to the target functionality: magnetic flux handling versus spin-dependent transport.

## 3.5. Annealed samples.

Samples grown at room temperature were also annealed at 500ºC to compare their properties with those of the samples grown on heated substrates. It was not possible to anneal above 500ºC because the sample suffered delamination due to mismatch in thermal expansion coefficient between substrate and film.

The behavior observed in X-ray diffraction is very different to that observed in samples grown on heated substrates. On both substrates, when annealed, the diffractogram remained practically the same as that observed at room temperature. No CMA peaks were seen in the case of MgO, while for Si only the (220) peak was present and remain visible up to 500ºC. No changes were observed in the case of resistivity either, remaining at its value of 260 μΩ·cm.

Despite all these differences, the evolution of the hysteresis loops is very similar to the series grown on heated substrates, with their squareness and saturation magnetization increasing with temperature and the coercive field decreasing. This indicates that, for applications where only saturation magnetization matters, annealing on Si samples is as good option as growing on heated substrates.

## 4. Conclusions

In summary, this work provides a detailed structural and magnetic characterization of 750 nm thick $Co_2MnAl$ films, establishing the critical processing windows for their integration on emerging technologies using MgO and Si substrates. Our experimental results lead to the following conclusions:

- Substrate-driven structural evolution. X-ray diffraction confirms that MgO (001) promotes a superior crystalline quality, with the B2 ordering degree increasing systematically with growth temperature up to 600°C. In contrast, films on Si (001) maintain structural stability only up to 500°C, beyond which severe interfacial degradation occurs.
- Samples grown on Si substrates exhibit a clear thermal limit at 500 °C. Above this temperature, Si diffuses rapidly into the Heusler layer, forming Co silicides. This results in Kirkendall voids, interfacial cavitation and microcracking, which explain both the abrupt rise in resistivity and a final disordered A2 phase. These features act as physical barriers to electronic transport, defining a clear thermal budget for CMOS-compatible integration.
- Magnetic robustness vs. Stoichiometry. Despite the Mn-poor composition ($Co_{2.25}Mn_{0.7}Al_{1.05}$) measured by EDX, the films exhibit a remarkably stable saturation magnetization of 600 emu/cm³. This demonstrates that the magnetic functionality of CMA remains robust against moderate off-stoichiometry, a key requirement for reliable device fabrication.
- Mechanical advantages of thick films. The 750 nm thickness investigated here proves to be a decisive factor in maintaining the structural integrity of the film. The SEM analysis shows that the film acts as a mechanical bridge over interfacial reaction zones, preventing immediate collapse despite the formation of Kirkendall voids.

These findings validate the use of thick $Co_2MnAl$ layers for applications requiring significant magnetic volume, such as integrated inductors or thermal energy harvesters, providing a clear guide for selecting growth parameters based on the chosen substrate

Acknowledgements:

This work was supported by the Spanish MICIN under PID2022-141373NB-I00 project, by the Junta de Comunidades de Castilla-La Mancha through "Proyecto de Investigacion Regional SBPLY/23/180225/000039", and by the "Plan Propio de la Universidad de Castilla-La Mancha" (FEDER, EU) 2022-GRIN-34313. The authors thank Mr. E. Prado for providing helpful studies using SEM technique.

CRediT authorship contribution statement:

David Robles-Cuenca: Data curation, Formal analysis, Investigation, Methodology, Validation, Visualization, Writing – original draft. Juan Antonio González: Conceptualization, Formal analysis, Methodology, Project administration, Resources, Supervision, Writing – review & editing. Juan Pedro Andrés: Conceptualization, Formal analysis, Methodology, Project administration, Resources, Supervision, Writing – review & editing. Miguel Ángel Arranz: Investigation, Visualization, Writing – review & editing. Mohamed Salaheldeen: Writing – review & editing. Ricardo López Antón: Conceptualization, Formal analysis, Methodology, Project administration, Resources, Supervision, Writing – review & editing.